\documentclass[
aps,
twocolumn,
longbibliography,
superscriptaddress,
floatfix,
]{revtex4} 

\usepackage{amsmath}
\usepackage{amssymb}
\usepackage{graphicx} 
\usepackage{color}
\usepackage{float}

\begin{document}
	
\title{Cracking urban mobility}
	
\author{H. A. Carmona} \affiliation{Departamento de F\'{i}sica,
Universidade Federal do Cear\'{a}, 60451-970 Fortaleza, Cear\'{a}, Brazil}
	
\author{A. W. T. de Noronha} \affiliation{Departamento de F\'{i}sica,
Universidade Federal do Cear\'{a}, 60451-970 Fortaleza, Cear\'{a}, Brazil}
	
\author{A. A. Moreira} \affiliation{Departamento de F\'{i}sica,
Universidade Federal do Cear\'{a}, 60451-970 Fortaleza, Cear\'{a}, Brazil}
	
\author{N. A. M. Ara\'{u}jo} \affiliation{Departamento de F\'{i}sica,
Universidade Federal do Cear\'{a}, 60451-970 Fortaleza, Cear\'{a}, Brazil}
\affiliation{Departamento de F\'{\i}sica, Faculdade de Ci\^{e}ncias,
Universidade de Lisboa, 1749-016 Lisboa, Portugal} \affiliation{Centro de
F\'{i}sica Te\'{o}rica e Computacional, Universidade de Lisboa, 1749-016
Lisboa, Portugal}
	
\author{J. S. Andrade Jr.} \email{soares@fisica.ufc.br}
\affiliation{Departamento de F\'{i}sica, Universidade Federal do Cear\'{a},
60451-970 Fortaleza, Cear\'{a}, Brazil}

\begin{abstract} 
Assessing the resilience of a road network is instrumental to improve existing
infrastructures and design new ones. Here we apply the optimal path crack model
(OPC) to investigate the mobility of road networks and propose a new proxy for
resilience of urban mobility. In contrast to static approaches, the OPC accounts
for the dynamics of rerouting as a response to traffic jams. Precisely, one
simulates a sequence of failures (cracks) at the most vulnerable segments of the
optimal origin-destination paths that are capable to collapse the system. Our
results with synthetic and real road networks reveal that their levels of
disorder, fractions of unidirectional segments and spatial correlations can
drastically affect the vulnerability to traffic congestion. By applying the OPC
to downtown Boston and Manhattan, we found that Boston is significantly more
vulnerable than Manhattan. This is compatible with the fact that Boston heads
the list of American metropolitan areas with the highest average time waste in
traffic. Moreover, our analysis discloses that the origin of this difference
comes from the intrinsic spatial correlations of each road network. Finally, we
argue that, due to their global influence, the most important cracks identified
with OPC can be used to pinpoint potential small rerouting and structural
changes in road networks that are capable to substantially improve urban
mobility.
\end{abstract}
	
\maketitle

\section{Introduction}
Traffic congestion is part of the daily life in a metropolitan region. In
the top 20 cities in the US, it is estimated that the average daily
commuter wastes more than 85 hours/year in traffic
congestion~\cite{GlobalTrafficReport2018}. Boston heads this list with a
160+ hours/year average delay. The numbers are even worst in cities like
Moscow, London, Bogota, or Mexico City, where the average time wasted per
year exceeds 200 hours. This inefficiency not only impacts on life quality
and the environment, but it also compromises economic growth. A recent
study using data from 88 US metropolitan areas suggests that a seemingly
harmless average delay of 4.5 minutes for each one-way auto commute in a
city is enough to slow down job growth~\cite{Sweet2014}.
	
To proper assess urban mobility, one needs to account for the impact of road
congestion in global
traffic~\cite{Louf2014,Colak2016,Sole-Ribalta2018,Barbosa2018,Barthelemy2019}.
Li~\textit{et al.}~\cite{Li2015} proposed to apply Percolation Theory to
evaluate how global connectivity is lost when vulnerable roads are congested.
Their static analysis for different hours of the day gives insight into normal
and rush-hour traffic and helped identifying vulnerable
roads~\cite{Gonzalez2008,Colak2016,Olmos2018,Zeng2019}.  However, in reality,
users are actively evaluating their routes and taking alternative paths to
avoid traffic jams~\cite{Zhu2015,Lima2016,Zhang2019}.  Thus, the probability
that a road gets congested depends, not only on its average level of traffic,
but also on the likelihood that users take it in their
route~\cite{Wang2012,Guo2019}. 
\begin{figure}
	\centering
	\includegraphics[width=0.6\columnwidth]{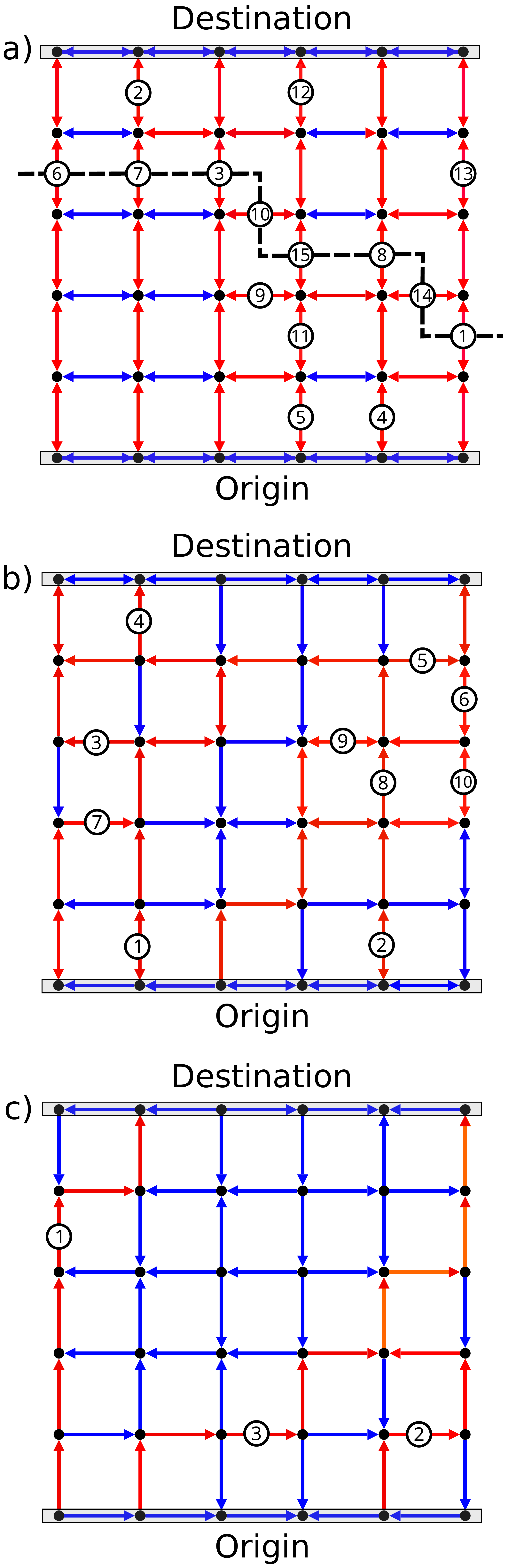}
	\caption{Sequence of removed links during the OPC process for $6\times 6$
		square lattices under weak disorder in travelling times ($\beta=0.002$) and
		different values of the fraction $p$ of unidirectional links, namely, (a)
		$p=0$, (b) $p=0.4$, and (c) $p=1$. Before the collapse of the system, all links
		in red were part of an optimal path, at least once, from the bottom (origin) to
		the top (destination) of the lattice. Those removed are indicated with white
		circles in the middle, numbered according to the OPC removal sequence, as
		explained in the main text. The number of removed links clearly decreases with
		$p$. The dashed line in (a) corresponds to the fracture backbone that is always
		present as a result of the OPC process applied to fully bidirectional networks
		($p=0$)~\cite{Andrade2009}.~\label{fig::spatial.distribution}}
	\end{figure}

\begin{figure}[t]
	\centering
	\includegraphics[width=\columnwidth]{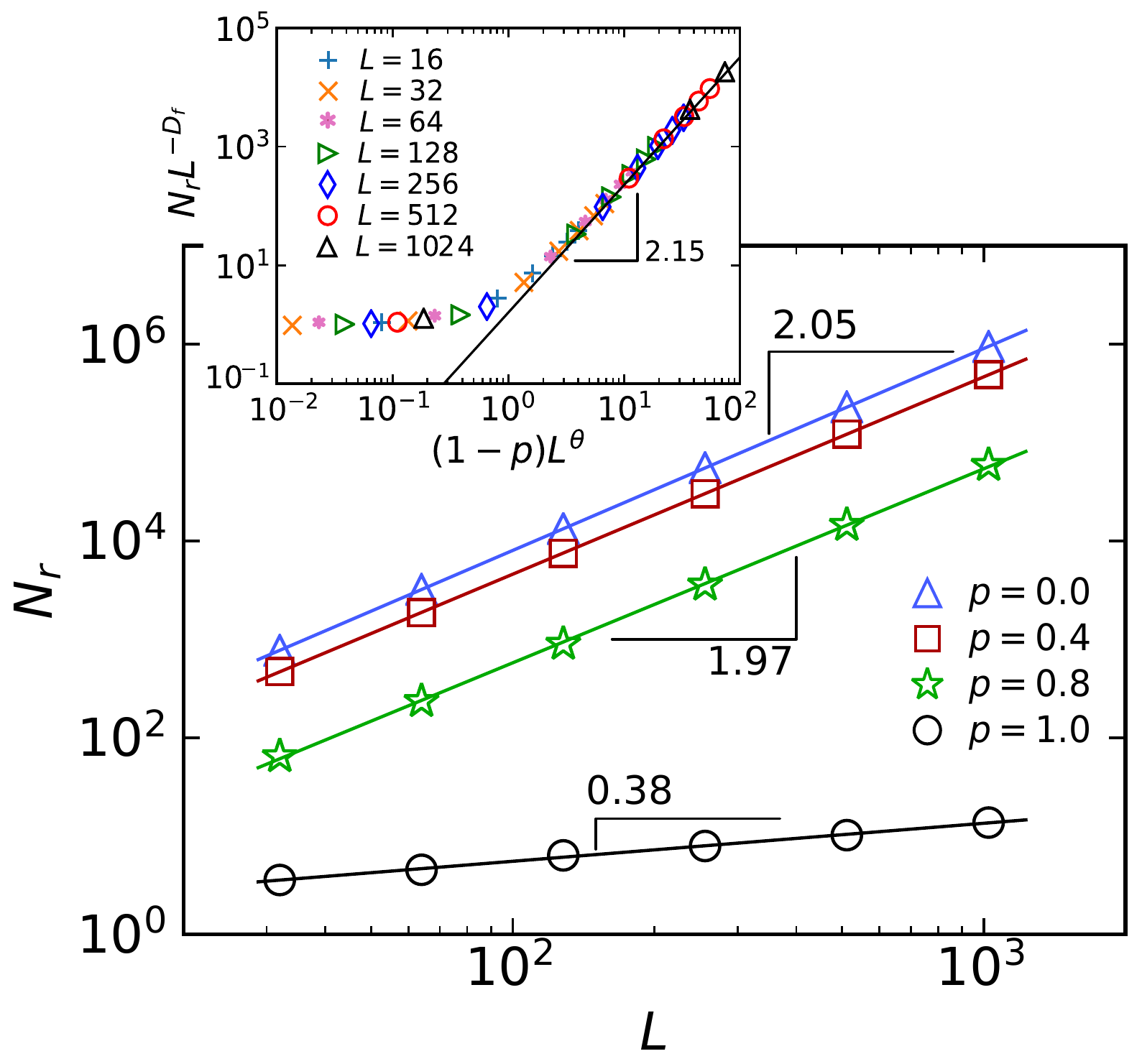}
\caption{Logarithmic dependence of the total number of removed links $N_r$ on
	the linear size of the lattice $L$ for different values of the fraction $p$ of
	unidirectional links. The symbols correspond to averages over $100$ thousand
	network realizations and weak disorder in their traveling times
	($\beta=0.002$). The results for $p=0.4$ and $0.8$ are consistent with the
	scaling, $N_r\sim L^2$, obtained for $p=0$, namely, fully bidirectional
	lattices~\cite{Andrade2009}. For a completely unidirectional lattice, $p=1$, we
	find that $N_r\sim L^{D_f}$, with $D_f=0.382\pm 0.002$. The error bars are
	smaller than the symbols. The inset shows the tricritical crossover scaling and
	data collapse for the OPC model on regular lattices. The scaling function is
	given by Eq.~\eqref{eq::ansatz}, with $D_f=0.38$ and $\theta=0.753$. The solid
	line in the inset represents a power law with exponent $2.15$.
	~\label{fig::number.removed.links}}
\end{figure}

\begin{figure}[t] 
	\centering
	\includegraphics[width=\columnwidth]{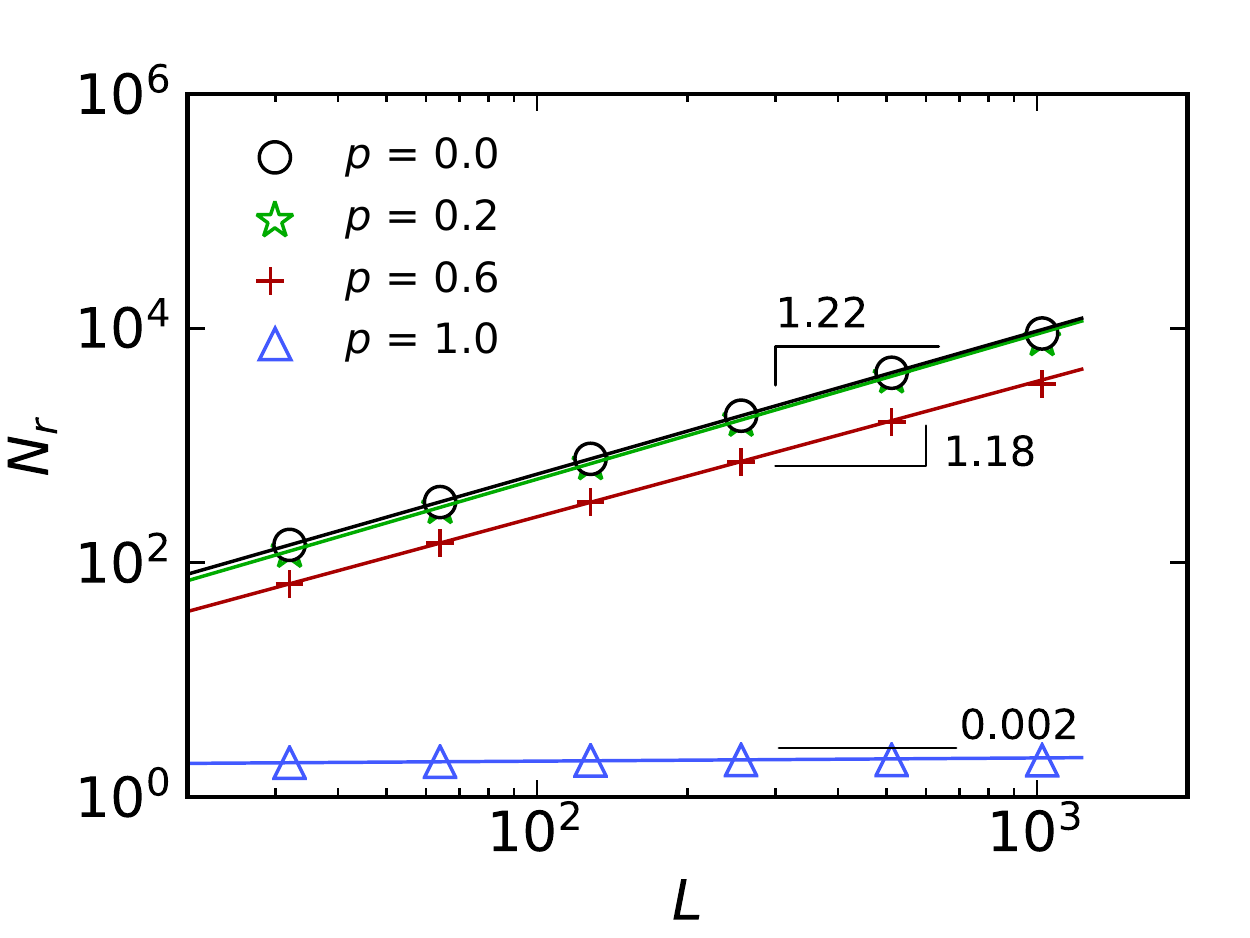} 
\caption{Logarithmic dependence of the total number of removed links $N_r$
on the linear size of the lattice $L$, for different values of the fraction
$p$ of unidirectional links. The symbols correspond to averages over $100$
thousand network realizations with sizes $L=16$, $32$, $64$, $512$, and
$1024$ and strong disorder in their traveling times ($\beta=400$). The
results for $p=0.4$ and $0.8$ are consistent with the scaling, $N_r\sim
L^D_f$, with $D_{f}=1.22\pm 0.01$ obtained for $p=0$, namely, fully
bidirectional lattices~\cite{Andrade2009}.  For a completely unidirectional lattice,
$p=1$, we find that $N_r\sim L^{D_f}$, with $D_f=0.002\pm 0.004$. The error
bars are smaller than the symbols.~\label{fig::s3}} 
\end{figure}
	
\begin{figure}[t]
	\centering
	\includegraphics[width=\columnwidth]{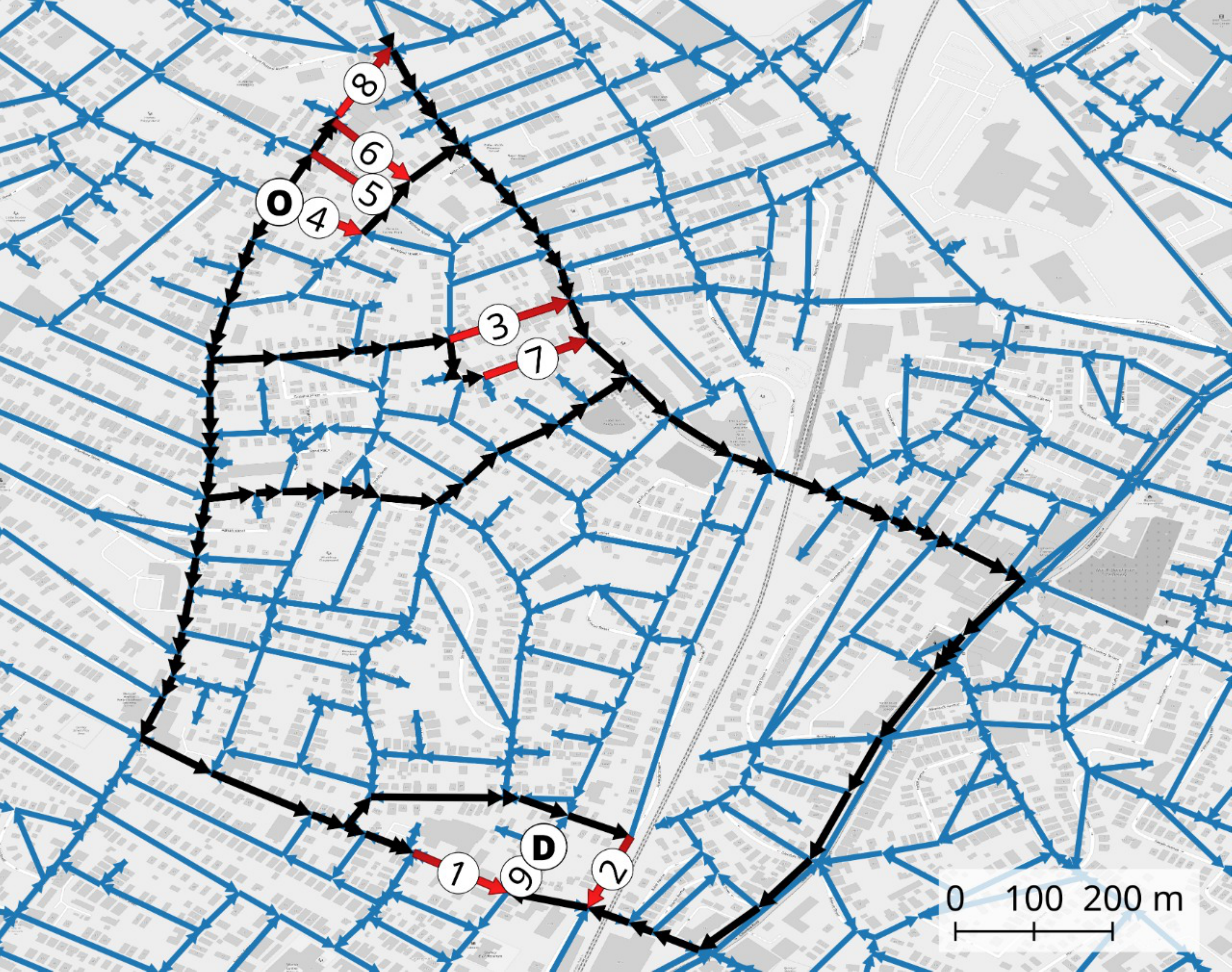} 
	\caption{Schematic representation of the OPC process for a realization
of a origin (O) destination (D) in downtown Boston.  Before the collapse
of the network, all links in black or red were part of an optimal path, at
least once. Those removed are in red and indicated with white circles in the
middle, numbered according to the OPC removal sequence, as explained in the
main text.
	~\label{fig::spatial.distribution.Boston}}
\end{figure}
	
Without central planning, travellers usually choose the route that minimizes
their travelling time. However, when a road segment gets congested, a new
optimal route needs to be found. The optimal path crack model (OPC) was
introduced as a general framework to study the resilience of a network
infrastructure to a sequence of optimal path
failures~\cite{Andrade2009,Oliveira2011}. The OPC is described as follows. Let
us consider a square lattice of size $L$ with periodic boundary conditions in
the horizontal direction and fixed boundary conditions at the top and bottom.
To each link, a travelling time $t$ is randomly assigned according to a given
probability distribution $P(t)$, with $t>0$. Using Dijkstra's
algorithm~\cite{Dijkstra1959}, the first optimal path is identified which
minimizes the total travelling time between the bottom and the top of the
lattice. We then search and remove the most vulnerable link along this path,
defined as the one with the highest travelling time. The next optimal path is
identified, which cannot contain the removed link, and its vulnerable link is
also removed. We proceed iteratively until the lattice is disrupted and no
more paths can be found. The OPC is then the set of all removed links. In the
limit of strong disorder, all cracks are located on a single self-similar
connected line of fractal dimension equal to $1.22$~\cite{Andrade2009}. As a
matter of fact, this exponent value is statistically identical to the fractal
dimension previously found for the optimal path line under strong
disorder~\cite{Cieplak1994, Cieplak1996, Porto1997, Porto1999}, ``strands'' in
Invasion Percolation~\cite{Cieplak1994, Cieplak1996}, paths on Minimum
Spanning Trees~\cite{Dobrin2001}, and watersheds on uncorrelated
landscapes~\cite{Oliveira2011,Schrenk2012}. In the case of weak disorder, the
cracks spread all over the entire network before global connectivity is lost,
so that the total number of removed links scales as $N_{r} \sim L^{d}$, where
$d$ is the topological dimension of the lattice~\cite{Andrade2009}.

So far, all studies on OPC considered non-directed networks. However, road
networks always have a non-negligible fraction of one-way roads, which is
expected to substantially affect the traffic
dynamics~\cite{Schwartz2002,Buzna2006,Sanchez2002}. In addition, by contrast to
previous static approaches, the fact that OPC accounts for the dynamics of
rerouting as a response to traffic jams suggests that this model may be a better
proxy for resilience of urban mobility. It is under this framework that here we
investigate the behavior of OPC when applied to synthetic and real road
networks. This paper is organized as follows. In Section II, we study the role
of disorder and unidirectionality on the OPC model applied to synthetic road
networks. In Section III, the OPC is applied to the real road networks of
Manhattan and downtown Boston. In Section IV, we present the conclusions of our
study.

\section{Role of disorder and unidirectionality on OPC}
In order to better understand the combined role of disorder and unidirectionality
on the OPC, we applied the model to synthetic road networks in which the links
are assigned to be unidirectional with probability $p$ and bidirectional with
probability $(1-p)$.  For $p=0$ we recover the OPC of a fully bidirectional
lattice, while for $p=1$ all links are unidirectional.  Disorder is introduced
by assigning the travelling times of unidirectional links according to a
hyperbolic distribution $p\left(\tau_i\right)\propto 1/\tau_i$, truncated
between $\tau_{max}=1$ and $\tau_{min} = \exp \left(-\beta\right)$, where $\beta
\geq 0$ is the parameter that controls the disorder.  Typical realizations of
the OPC model for small networks generated with weak disorder are shown in
Fig.~\ref{fig::spatial.distribution}. Note that each bidirectional link in the
network has distinct travelling times associated to its two directions.  As
previously observed~\cite{Andrade2009}, the set of OPC cracks generated in fully
bidirectional networks ($p=0$) always contains a contiguous subset that spans
the entire system from left to right, regardless of the level of disorder. In
the presence of any amount of unidirectional links ($p>0$), however, the cracks
do not necessarily form a contiguous fracture that divides the network into two
pieces. Moreover, the larger the value of $p$, rarer is the occurrence of this
spanning fracture. When the unidirectional links are assigned independently at
random, there is always global connectivity (percolation), but sink nodes and
closed loops are formed~\cite{Noronha2018}.  The links belonging to such closed
loop will never be part of a shortest path and thus the fracture does not need
to be contiguous, as they also contribute to separate the lattice into two
pieces.

Figure~\ref{fig::number.removed.links} shows the logarithmic dependence of the
number of removed links $N_r$ on the linear size of the lattice $L$, for
lattices with sizes varying in the range $16\leq L\leq 512$ and weak disorder
in their distribution of travelling times ($\beta=0.002$).  For 
$p\neq 1$, the numerical results are consistent with $N_r\sim L^d$, where
$d=2$. Thus, provided that a non-zero fraction of links is bidirectional, the
set of all removed links is compact, as reported for networks with only
bidirectional links ($p=0$)~\cite{Andrade2009}.  Nevertheless, the fraction of
removed links is a monotonic decreasing function of $p$, meaning that the
prefactor of the power-law relation decreases too (see
Fig.~\ref{fig::number.removed.links}). Surprisingly, in the limit case
of a completely unidirectional lattice, $p=1$, we find that $N_r\sim
L^{D_f}$, with $D_f=0.382\pm 0.002$. Our result
therefore indicates that the OPC set at this point belongs to a different
universality class. Moreover, since $D_f<d$, the fraction of removed links for
an infinite lattice (thermodynamic limit) is zero for $p=1$.

The statistics of the OPC set generated under weak disorder suggests that its
dimension does crossover from $d$, for $p<1$, to $D_f$ at $p=p_c=1$. This crossover
is analogous to what is observed at the theta point of polymer
systems~\cite{deGennes1979,Chang1991,Poole1989}. At high temperatures, the
configurations of a polymer chain are well described by a self-avoiding random
walk, as the only relevant interactions are excluded volume. However, at the
theta-temperature, the attractive forces are no longer negligible, and the
statistics are then different. Also, in ranked surfaces, when occupying links
sequentially, but suppressing global connectivity, the fractal dimension of the
set of links that are not occupied due to this constraint changes from $3/4$ at
the percolation threshold to $1.22$ above it~\cite{Schrenk2012}. For the case of
OPC, we consider the following crossover \textit{Ansatz}:
\begin{equation}
	\label{eq::ansatz}
	N_r=L^{D_f}\mathcal{F}\left[\left(p_c-p\right)L^\theta\right],
\end{equation} 
where $\theta$ is the crossover exponent, $\mathcal{F}\left[x\right]\sim
x^{\eta}$ for $x\neq0$ and equal to a nonzero constant at $x=0$. The inset
in Fig.~\ref{fig::number.removed.links} shows the data collapse obtained
with this tricritical scaling for different lattice sizes and values of
$p$, with $\theta=0.753$. By fitting the power-law regime of the scaling
function $\mathcal{F}$, we estimate $\eta=2.15\pm0.03$. From the
\textit{Ansatz}, we expect that, 
\begin{equation}\label{eq::exponents} %
	D_f+\eta\theta=d, %
\end{equation} 
what is verified, within error bars. This behavior confirms that the
universality class of OPC is robust and only breaks down for a fully
unidirectional lattice ($p=1$).

For the case of local travelling times
with strong disorder, $\beta=400$ (see Fig.~\ref{fig::s3}), the OPC
results for $p\neq 1$ are also consistent with the behavior previously
observed for fully bidirectional networks ($p=0$), namely, $D_{f}=1.22\pm
0.01$~\cite{Andrade2009}. As in the weak disorder case, it is only for $p=1$
that the finite-size scaling of the system becomes noticeably different, with
$D_f=0.002\pm 0.004$. A rather small number of removed links is therefore
sufficient to block a fully directed lattice subjected to strong disorder.
Summing up, our analysis with synthetic road networks shows that the higher
the fraction of unidirectional segments and the level of disorder are, the
lower is the resilience for urban mobility. 

\begin{figure}[ht]
	\centering
	\includegraphics[width=\columnwidth]{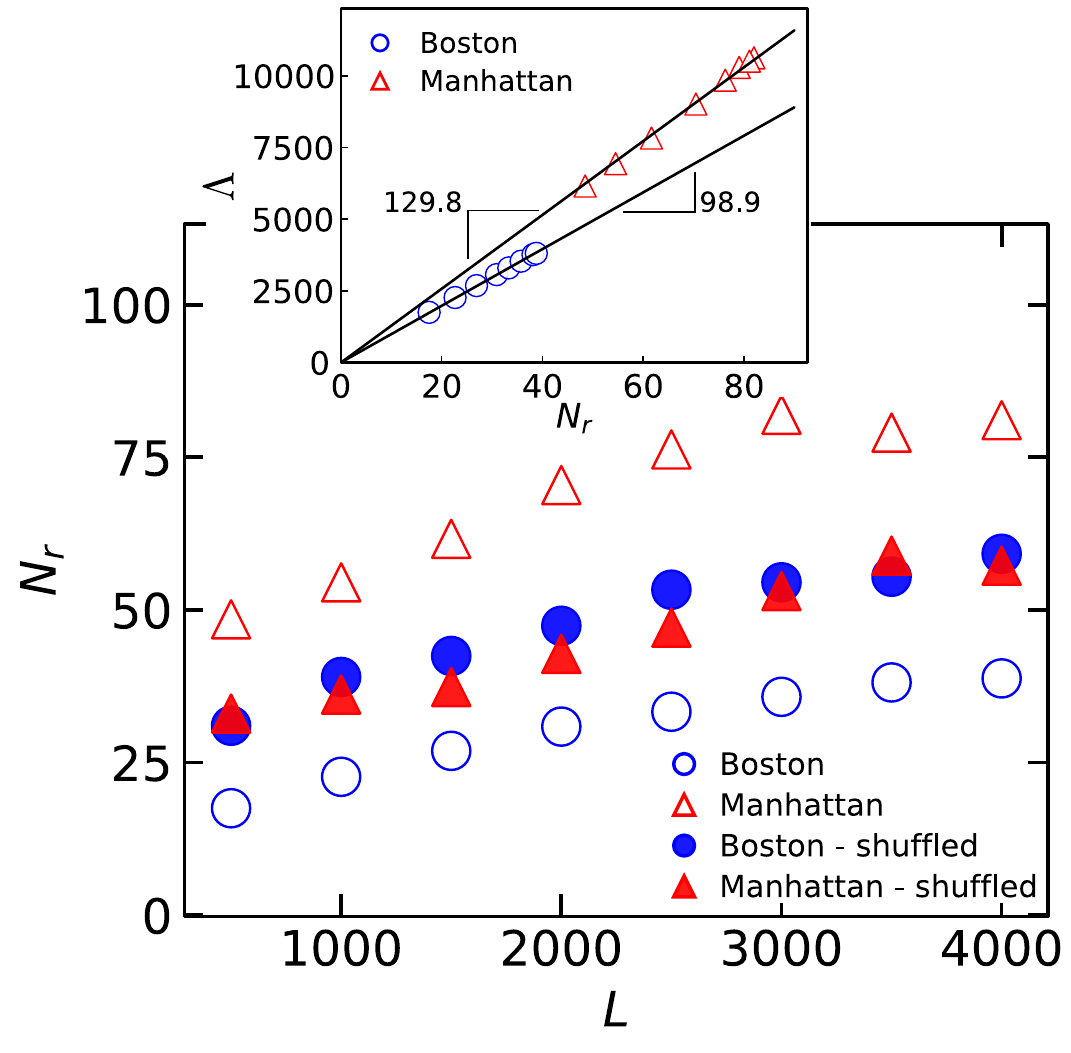} 
    \caption{Dependence of the total number of removed roads $N_r$ on the
	origin-destination distance $L$ (in meters) for downtown Boston and
	Manhattan (open symbols). Numerical results are averages over 2000 OD
	samples for each city. Also shown  are the results of OPC simulations
	preserving the geometry of the road networks, but shuffling the values of
	$t/\ell$ among randomly chosen pairs of road segments (filled symbols). The
	inset shows the linear dependence, $\Lambda=a N_r$, with $a=98.9 \pm 0.03$
	and $129.8\pm 0.04$ for Boston and Manhattan, respectively. In all cases,
	the error bars are smaller than the symbols.~\label{fig::real.data}}
\end{figure} 

\begin{figure}[t]
	\centering
	\includegraphics[width=\columnwidth]{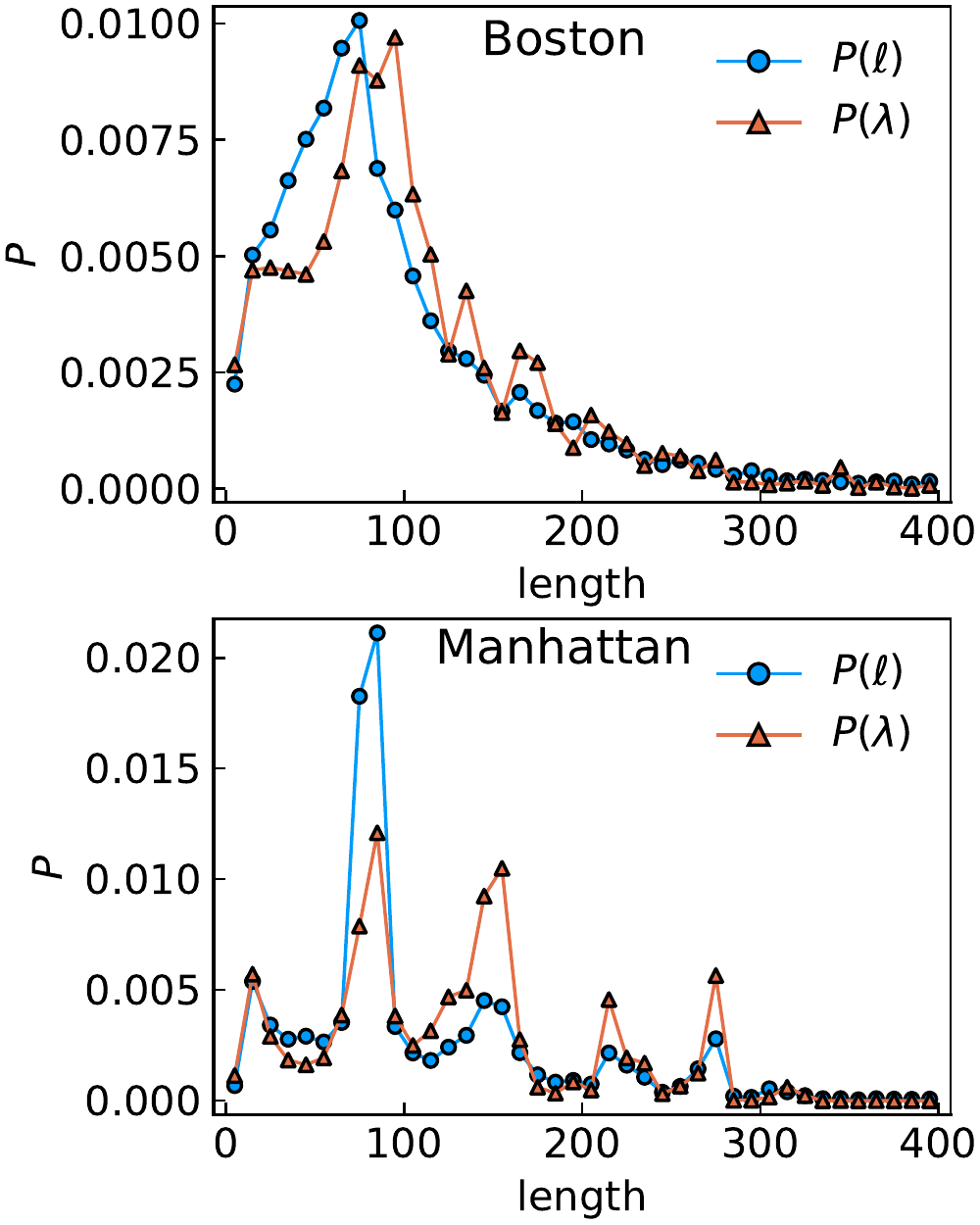} 
\caption{ Comparison between distributions of the lengths $\ell$ of all
road segments, and the lengths $\lambda$ of those among all road segments
that have been removed  during the OPC process applied to Boston (top) and
Manhattan (bottom).\label{fig::distributions.ell.lambda}}
\end{figure}

\begin{figure}[t]
	\centering
	\includegraphics[width=\columnwidth]{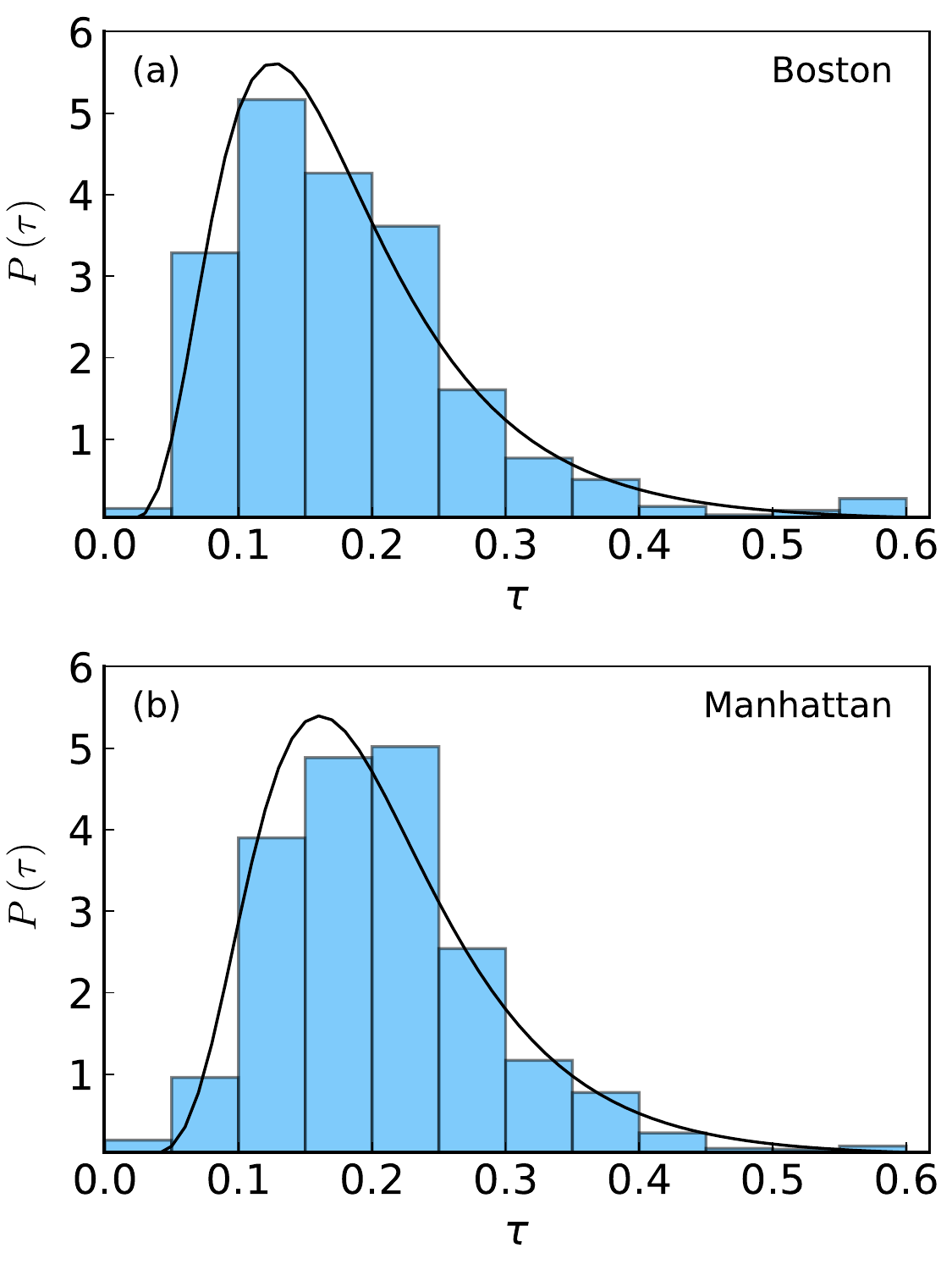}
	\caption{Distribution $P(\tau)$ of the travelling time per length
$\tau=t/\ell$ of the road segment for Boston (a) and Manhattan (b). The
solid lines are best fits to a lognormal distribution with mean $0.18$ and
variance $0.1$ for Boston, and $0.21$ and $0.09$ for Manhattan.~\label{fig::disorder}
	}
\end{figure}

\begin{figure*}[ht] 
	\centering
	\includegraphics[width=\textwidth]{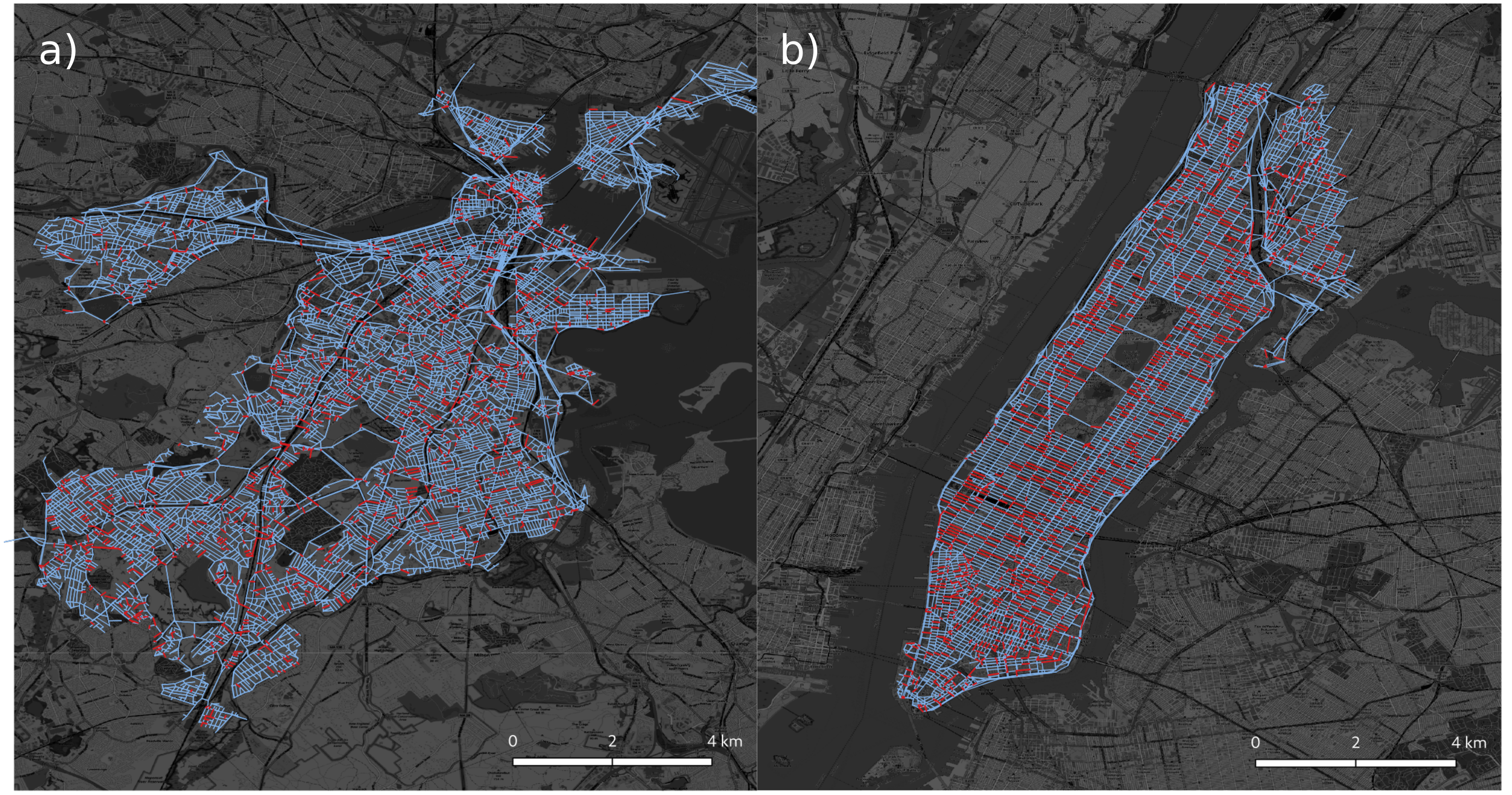}
	\caption{Road maps for downtown Boston (left) and Manhattan (right). The thick
	red lines are the road segments that have been removed first. All other road
	segments are shown in blue. Only OPC samples performed with OD distances
	$L=2000$ m have been considered.~\label{fig::real.data.maps}}
\end{figure*} %

\begin{figure}[t]
\centering
\includegraphics[width=\columnwidth]{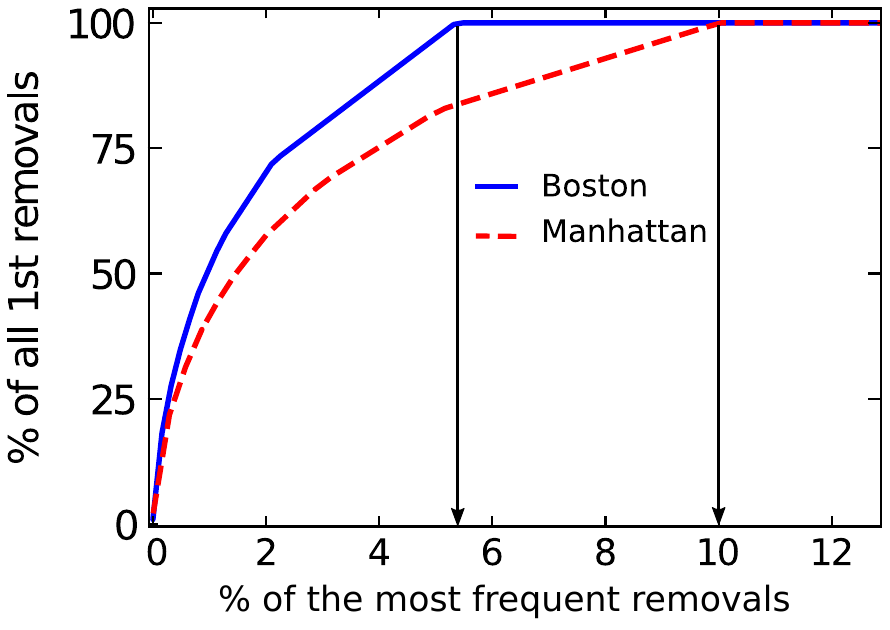} 
\caption{Cumulative dependence of the percentage of all first removals
during the OPC process on the fraction of the most frequent ones. The
vertical solid lines indicate that the first removals correspond to $5.4\%$
of all road segments in the case of Boston, while $10.0\%$ is the
percentage required for Manhattan.~\label{fig::cummutative.fraction.removed}}
\end{figure}

\section{OPC for real road networks}
     
Next, we show that the OPC can be effectively used as a proxy for urban mobility. 
In general, the fraction of one-way roads changes from city to city. Here we
apply the OPC method to two metropolitan areas in US, namely, Manhattan and
downtown Boston, the latter being the top one in the rank of North America for
average time waste in traffic~\cite{GlobalTrafficReport2018}. For this purpose,
we obtained the road network structures of both urban areas from
OpenStreetMap~\cite{OpenStreetMap}. Their corresponding average travelling times
per road were downloaded from Directions API Google~\cite{GoogleDirections},
specifically for November 21th of 2018 at 4:00 am. This off-peak data set has
been chosen because, by construction, the travelling times should somehow
reflect the freeway-state of the streets and avenues constituting a given urban
mobility system. In order to assess the efficiency of these road networks to
urban traffic, the OPC is numerically calculated for different pairs of
origin-destination (OD) sites. Precisely, for each OPC realization, one site of
the network is selected at random to be the center of a circle of radius $L$. A
point on this circle is then randomly chosen and the closest road sites to the
center and to this point are taken as the origin and destination, respectively,
only if the distance between them is equal to $L$ within a tolerance of $5\%$.
Here, due to the fact that road segments composing the urban networks have
different travelling times $t$ as well as lengths $\ell$, an optimal path is
identified among all possible paths during the OPC process as the one with the
minimum sum of $t/\ell$ over all its road segments.
	
OPC simulations have then been performed with $2000$ OD realizations for
Boston and Manhattan. A typical realization of the OPC for downtown Boston is
shown in Fig.~\ref{fig::spatial.distribution.Boston}.  One should note that the
bidirectional links (two-way streets and avenues) are in fact composed of two
unidirectional ones of contrary directions that can be removed independently
during the OPC process.  In Fig.~\ref{fig::real.data}, we show how the average
number of removed sites $N_r$ varies with the distance $L$ for both urban
areas. The smaller $N_r$ is, less resilient is the road network. These results
indicate that Boston is systematically much more vulnerable than Manhattan,
regardless of the origin-destination distance $L$, what is consistent with
their relative positions in the national rank of urban mobility. Moreover, the
same behavior can be observed if, instead of $N_r$, we quantify the result of
each OPC process in terms of the sum $\Lambda=\sum_{i=1}^{N_r}\lambda_i$,
where $\lambda_i$ is equal to the length $\ell$ of the removed road segment at
iteration $i$.  Interestingly, as shown in the inset of
Fig.~\ref{fig::real.data}, these two measures are, in fact, linearly related
$\left<\Lambda\right>=a N_r$, with $a = 98.9\,\rm{m}$ and $129.8\,\rm{m}$ for
Boston and Manhattan, respectively. For all practical purposes, this indicates
the absence of spatial statistical correlations in the OPC process of
selecting $\lambda$ from the distribution of segment lengths $P(\ell)$ of both
road networks. As shown in Fig.~\ref{fig::distributions.ell.lambda}, the
distributions $P(\ell)$ and $P(\lambda)$ for a given road network display
similar qualitative features in shape. More important, these results for
Boston and Manhattan indicate that the variable $\lambda$ is selected by the
OPC process over the entire range of $\ell$ values. One should note that,
although the OPC process tends to pick larger values of $\lambda$ from
$P(\ell)$ more frequently, this does not imply the presence of statistical
correlations among links in the road networks. Such a behavior clearly
reflects a local feature of the network that is captured by the OPC but
cannot explain the substantial differences in the resilience of Boston and
Manhattan, since the OPC results for both road networks follow the same
tendency.

We now investigate the origin of the substantial difference between the
resilience of Boston and Manhattan. Figure~\ref{fig::disorder} shows the
distributions of travelling time per segment length $P(t/\ell)$ obtained for
their corresponding road networks. Clearly, the disorder is weak in both cases
and the modest difference between them is not compatible with the large
discrepancy observed in their resilience. Furthermore, the fraction of
unidirectional links $p$ found for downtown Boston is $0.65$, while the value
for Manhattan is $0.78$.  However, our results from synthetic road networks
indicate that, for similar levels of disorder, the resilience of the system
should decrease with the value of $p$. As
a consequence, the observed difference in the resilience of Boston and Manhattan
also cannot stem from their distinct $p$ values.  One therefore can only rely on
the particular features of the intrinsic spatial correlations that should be
present in these urban systems to justify their very distinct responses. In
order to test for this hypothesis, we performed additional OPC simulations
preserving the geometry of both networks, but shuffling the values of $t/\ell$
among randomly chosen pairs of road segments. The results presented in
Fig.~\ref{fig::real.data} are rather surprising and twofold.  First, they show
that the effect of suppressing the spatial correlations is to practically
collapse the curves $N_r$ against $L$ of both cities to a single curve,
therefore demonstrating the generality and comparative power of the OPC method
for urban mobility. Second, the fact that the values of $N_r$ systematically
increase for Boston and decrease for Manhattan, as compared to the results using
the real (non-shuffled) data sets, corroborates the ability of our approach to
properly capture negative and positive effects of spatial correlations on urban
mobility.
	
One question that naturally arises is how the OPC method can be used to
enhance urban mobility. A possible answer is to prioritize rerouting and
structural improvements based on the identification of those road segments
that are more frequently appearing among the first removals in all OPC
sequences. In Fig.~\ref{fig::real.data.maps} are the road maps for the two
urban areas, where the highlighted road segments (thick and red) correspond to
all those removed first, considering all OPC samples performed with OD
distances $L=2000$ m. For downtown Boston (see
Fig.~\ref{fig::real.data.maps}a), we find that only a percentage of $5.4\%$ of
all road segments is removed, while $10.0\%$ is the value for Manhattan (see
Fig.~\ref{fig::real.data.maps}b). The cumulative dependence of the percentage
of all first removals on the percentage of the most frequent ones is shown in
Fig.~\ref{fig::cummutative.fraction.removed}. These results suggest that, as
compared to Manhattan, a relatively small number of potential local changes in
Boston might be very efficient in improving urban mobility. The fact that
Manhattan is much more resilient should therefore increase the costs of
improvements.

\section{Conclusions}
In summary, we proposed a proxy for resilience of urban mobility based on OPC.
Our results with synthetic and real road networks suggest that their
vulnerability to traffic congestion are strongly dependent on the level of
disorder, fraction of unidirectional segments and intrinsic spatial
correlations. These observations have practical implications in the design and
restructuring for improved urban mobility. With OPC, we obtain the list of
most vulnerable links, defined as the ones with the highest travelling time
along the sequence of shortest origin-destination paths. The vulnerability of
such a link might stem from many design factors, such as, the capacity of the
road, speed limit, number of crossroads or traffic-light dynamics. We conclude
that OPC is a general and powerful method to access urban mobility, and gives
practical insight that can effectively help identifying and mitigating
vulnerabilities of real road networks.

\begin{acknowledgments}  
We gratefully acknowledge financial support from the Brazilian agencies CNPq, CAPES and
FUNCAP, the National Institute of Science and Technology for Complex Systems in
Brazil, and from the Portuguese Foundation for Science and Technology (FCT)
under Contracts no. UIDB/00618/2020 and UIDP/00618/2020.%
\end{acknowledgments}

%\bibliography{references.bib}

\begin{thebibliography}{34}
	\expandafter\ifx\csname natexlab\endcsname\relax\def\natexlab#1{#1}\fi
	\expandafter\ifx\csname bibnamefont\endcsname\relax
	  \def\bibnamefont#1{#1}\fi
	\expandafter\ifx\csname bibfnamefont\endcsname\relax
	  \def\bibfnamefont#1{#1}\fi
	\expandafter\ifx\csname citenamefont\endcsname\relax
	  \def\citenamefont#1{#1}\fi
	\expandafter\ifx\csname url\endcsname\relax
	  \def\url#1{\texttt{#1}}\fi
	\expandafter\ifx\csname urlprefix\endcsname\relax\def\urlprefix{URL }\fi
	\providecommand{\bibinfo}[2]{#2}
	\providecommand{\eprint}[2][]{\url{#2}}
	
	\bibitem[{\citenamefont{Reed and Kidd}(2018)}]{GlobalTrafficReport2018}
	\bibinfo{author}{\bibfnamefont{T.}~\bibnamefont{Reed}} \bibnamefont{and}
	  \bibinfo{author}{\bibfnamefont{J.}~\bibnamefont{Kidd}}, \bibinfo{type}{Tech.
	  Rep.}, \bibinfo{institution}{IRIX Research, United States}
	  (\bibinfo{year}{2018}).
	
	\bibitem[{\citenamefont{Sweet}(2014)}]{Sweet2014}
	\bibinfo{author}{\bibfnamefont{M.}~\bibnamefont{Sweet}},
	  \bibinfo{journal}{Urban Studies} \textbf{\bibinfo{volume}{51}},
	  \bibinfo{pages}{2088} (\bibinfo{year}{2014}).
	
	\bibitem[{\citenamefont{Louf and Barthelemy}(2014)}]{Louf2014}
	\bibinfo{author}{\bibfnamefont{R.}~\bibnamefont{Louf}} \bibnamefont{and}
	  \bibinfo{author}{\bibfnamefont{M.}~\bibnamefont{Barthelemy}},
	  \bibinfo{journal}{Sci. Rep.} \textbf{\bibinfo{volume}{4}}, \bibinfo{pages}{1}
	  (\bibinfo{year}{2014}).
	
	\bibitem[{\citenamefont{Colak et~al.}(2016)\citenamefont{Colak, Lima, and
	  Gonz{\'{a}}lez}}]{Colak2016}
	\bibinfo{author}{\bibfnamefont{S.}~\bibnamefont{Colak}},
	  \bibinfo{author}{\bibfnamefont{A.}~\bibnamefont{Lima}}, \bibnamefont{and}
	  \bibinfo{author}{\bibfnamefont{M.~C.} \bibnamefont{Gonz{\'{a}}lez}},
	  \bibinfo{journal}{Nat. Commun.} \textbf{\bibinfo{volume}{7}},
	  \bibinfo{pages}{10793} (\bibinfo{year}{2016}).
	
	\bibitem[{\citenamefont{Sol{\'{e}}-Ribalta
	  et~al.}(2018)\citenamefont{Sol{\'{e}}-Ribalta, G{\'{o}}mez, and
	  Arenas}}]{Sole-Ribalta2018}
	\bibinfo{author}{\bibfnamefont{A.}~\bibnamefont{Sol{\'{e}}-Ribalta}},
	  \bibinfo{author}{\bibfnamefont{S.}~\bibnamefont{G{\'{o}}mez}},
	  \bibnamefont{and} \bibinfo{author}{\bibfnamefont{A.}~\bibnamefont{Arenas}},
	  \bibinfo{journal}{Networks Spat. Econ.} \textbf{\bibinfo{volume}{18}},
	  \bibinfo{pages}{33} (\bibinfo{year}{2018}).
	
	\bibitem[{\citenamefont{Barbosa et~al.}(2018)\citenamefont{Barbosa, Barthelemy,
	  Ghoshal, James, Lenormand, Louail, Menezes, Ramasco, Simini, and
	  Tomasini}}]{Barbosa2018}
	\bibinfo{author}{\bibfnamefont{H.}~\bibnamefont{Barbosa}},
	  \bibinfo{author}{\bibfnamefont{M.}~\bibnamefont{Barthelemy}},
	  \bibinfo{author}{\bibfnamefont{G.}~\bibnamefont{Ghoshal}},
	  \bibinfo{author}{\bibfnamefont{C.~R.} \bibnamefont{James}},
	  \bibinfo{author}{\bibfnamefont{M.}~\bibnamefont{Lenormand}},
	  \bibinfo{author}{\bibfnamefont{T.}~\bibnamefont{Louail}},
	  \bibinfo{author}{\bibfnamefont{R.}~\bibnamefont{Menezes}},
	  \bibinfo{author}{\bibfnamefont{J.~J.} \bibnamefont{Ramasco}},
	  \bibinfo{author}{\bibfnamefont{F.}~\bibnamefont{Simini}}, \bibnamefont{and}
	  \bibinfo{author}{\bibfnamefont{M.}~\bibnamefont{Tomasini}},
	  \bibinfo{journal}{Phys. Rep.} \textbf{\bibinfo{volume}{734}},
	  \bibinfo{pages}{1} (\bibinfo{year}{2018}).
	
	\bibitem[{\citenamefont{Barthelemy}(2019)}]{Barthelemy2019}
	\bibinfo{author}{\bibfnamefont{M.}~\bibnamefont{Barthelemy}},
	  \bibinfo{journal}{Nature Reviews Physics} \textbf{\bibinfo{volume}{1}},
	  \bibinfo{pages}{406} (\bibinfo{year}{2019}).
	
	\bibitem[{\citenamefont{Li et~al.}(2015)\citenamefont{Li, Fu, Wang, Lu,
	  Berezin, Stanley, and Havlin}}]{Li2015}
	\bibinfo{author}{\bibfnamefont{D.}~\bibnamefont{Li}},
	  \bibinfo{author}{\bibfnamefont{B.}~\bibnamefont{Fu}},
	  \bibinfo{author}{\bibfnamefont{Y.}~\bibnamefont{Wang}},
	  \bibinfo{author}{\bibfnamefont{G.}~\bibnamefont{Lu}},
	  \bibinfo{author}{\bibfnamefont{Y.}~\bibnamefont{Berezin}},
	  \bibinfo{author}{\bibfnamefont{H.~E.} \bibnamefont{Stanley}},
	  \bibnamefont{and} \bibinfo{author}{\bibfnamefont{S.}~\bibnamefont{Havlin}},
	  \bibinfo{journal}{Proc. Natl. Acad. Sci. U. S. A.}
	  \textbf{\bibinfo{volume}{112}}, \bibinfo{pages}{669} (\bibinfo{year}{2015}).
	
	\bibitem[{\citenamefont{Gonz{\'{a}}lez
	  et~al.}(2008)\citenamefont{Gonz{\'{a}}lez, Hidalgo, and
	  Barab{\'{a}}si}}]{Gonzalez2008}
	\bibinfo{author}{\bibfnamefont{M.~C.} \bibnamefont{Gonz{\'{a}}lez}},
	  \bibinfo{author}{\bibfnamefont{C.~A.} \bibnamefont{Hidalgo}},
	  \bibnamefont{and} \bibinfo{author}{\bibfnamefont{A.-L.}
	  \bibnamefont{Barab{\'{a}}si}}, \bibinfo{journal}{Nature}
	  \textbf{\bibinfo{volume}{453}}, \bibinfo{pages}{779} (\bibinfo{year}{2008}).
	
	\bibitem[{\citenamefont{Olmos et~al.}(2018)\citenamefont{Olmos, {\c{C}}olak,
	  Shafiei, Saberi, and Gonz{\'{a}}lez}}]{Olmos2018}
	\bibinfo{author}{\bibfnamefont{L.~E.} \bibnamefont{Olmos}},
	  \bibinfo{author}{\bibfnamefont{S.}~\bibnamefont{{\c{C}}olak}},
	  \bibinfo{author}{\bibfnamefont{S.}~\bibnamefont{Shafiei}},
	  \bibinfo{author}{\bibfnamefont{M.}~\bibnamefont{Saberi}}, \bibnamefont{and}
	  \bibinfo{author}{\bibfnamefont{M.~C.} \bibnamefont{Gonz{\'{a}}lez}},
	  \bibinfo{journal}{Proc. Natl. Acad. Sci. U. S. A.}
	  \textbf{\bibinfo{volume}{115}}, \bibinfo{pages}{12654}
	  (\bibinfo{year}{2018}).
	
	\bibitem[{\citenamefont{Zeng et~al.}(2019)\citenamefont{Zeng, Li, Guo, Gao,
	  Gao, {Eugene Stanley}, and Havlin}}]{Zeng2019}
	\bibinfo{author}{\bibfnamefont{G.}~\bibnamefont{Zeng}},
	  \bibinfo{author}{\bibfnamefont{D.}~\bibnamefont{Li}},
	  \bibinfo{author}{\bibfnamefont{S.}~\bibnamefont{Guo}},
	  \bibinfo{author}{\bibfnamefont{L.}~\bibnamefont{Gao}},
	  \bibinfo{author}{\bibfnamefont{Z.}~\bibnamefont{Gao}},
	  \bibinfo{author}{\bibfnamefont{H.}~\bibnamefont{{Eugene Stanley}}},
	  \bibnamefont{and} \bibinfo{author}{\bibfnamefont{S.}~\bibnamefont{Havlin}},
	  \bibinfo{journal}{Proc. Natl. Acad. Sci. U. S. A.}
	  \textbf{\bibinfo{volume}{116}}, \bibinfo{pages}{23} (\bibinfo{year}{2019}).
	
	\bibitem[{\citenamefont{Zhu and Levinson}(2015)}]{Zhu2015}
	\bibinfo{author}{\bibfnamefont{S.}~\bibnamefont{Zhu}} \bibnamefont{and}
	  \bibinfo{author}{\bibfnamefont{D.}~\bibnamefont{Levinson}},
	  \bibinfo{journal}{PLoS One} \textbf{\bibinfo{volume}{10}},
	  \bibinfo{pages}{e0134322} (\bibinfo{year}{2015}).
	
	\bibitem[{\citenamefont{Lima et~al.}(2016)\citenamefont{Lima, Stanojevic,
	  Papagiannaki, Rodriguez, and Gonzalez}}]{Lima2016}
	\bibinfo{author}{\bibfnamefont{A.}~\bibnamefont{Lima}},
	  \bibinfo{author}{\bibfnamefont{R.}~\bibnamefont{Stanojevic}},
	  \bibinfo{author}{\bibfnamefont{D.}~\bibnamefont{Papagiannaki}},
	  \bibinfo{author}{\bibfnamefont{P.}~\bibnamefont{Rodriguez}},
	  \bibnamefont{and} \bibinfo{author}{\bibfnamefont{M.~C.}
	  \bibnamefont{Gonzalez}}, \bibinfo{journal}{J. R. Soc. Interface}
	  \textbf{\bibinfo{volume}{13}} (\bibinfo{year}{2016}).
	
	\bibitem[{\citenamefont{Zhang et~al.}(2019)\citenamefont{Zhang, Zeng, Li,
	  Huang, {Eugene Stanley}, and Havlin}}]{Zhang2019}
	\bibinfo{author}{\bibfnamefont{L.}~\bibnamefont{Zhang}},
	  \bibinfo{author}{\bibfnamefont{G.}~\bibnamefont{Zeng}},
	  \bibinfo{author}{\bibfnamefont{D.}~\bibnamefont{Li}},
	  \bibinfo{author}{\bibfnamefont{H.~J.} \bibnamefont{Huang}},
	  \bibinfo{author}{\bibfnamefont{H.}~\bibnamefont{{Eugene Stanley}}},
	  \bibnamefont{and} \bibinfo{author}{\bibfnamefont{S.}~\bibnamefont{Havlin}},
	  \bibinfo{journal}{Proc. Natl. Acad. Sci. U. S. A.}
	  \textbf{\bibinfo{volume}{116}}, \bibinfo{pages}{8673} (\bibinfo{year}{2019}).
	
	\bibitem[{\citenamefont{Wang et~al.}(2012)\citenamefont{Wang, Hunter, Bayen,
	  Schechtner, and Gonz{\'{a}}lez}}]{Wang2012}
	\bibinfo{author}{\bibfnamefont{P.}~\bibnamefont{Wang}},
	  \bibinfo{author}{\bibfnamefont{T.}~\bibnamefont{Hunter}},
	  \bibinfo{author}{\bibfnamefont{A.~M.} \bibnamefont{Bayen}},
	  \bibinfo{author}{\bibfnamefont{K.}~\bibnamefont{Schechtner}},
	  \bibnamefont{and} \bibinfo{author}{\bibfnamefont{M.~C.}
	  \bibnamefont{Gonz{\'{a}}lez}}, \bibinfo{journal}{Sci. Rep.}
	  \textbf{\bibinfo{volume}{2}}, \bibinfo{pages}{1001} (\bibinfo{year}{2012}).
	
	\bibitem[{\citenamefont{Guo et~al.}(2019)\citenamefont{Guo, Zhou, Fan, Tong,
	  Zhu, Lv, Li, and Havlin}}]{Guo2019}
	\bibinfo{author}{\bibfnamefont{S.}~\bibnamefont{Guo}},
	  \bibinfo{author}{\bibfnamefont{D.}~\bibnamefont{Zhou}},
	  \bibinfo{author}{\bibfnamefont{J.}~\bibnamefont{Fan}},
	  \bibinfo{author}{\bibfnamefont{Q.}~\bibnamefont{Tong}},
	  \bibinfo{author}{\bibfnamefont{T.}~\bibnamefont{Zhu}},
	  \bibinfo{author}{\bibfnamefont{W.}~\bibnamefont{Lv}},
	  \bibinfo{author}{\bibfnamefont{D.}~\bibnamefont{Li}}, \bibnamefont{and}
	  \bibinfo{author}{\bibfnamefont{S.}~\bibnamefont{Havlin}},
	  \bibinfo{journal}{EPJ Data Sci.} \textbf{\bibinfo{volume}{8}},
	  \bibinfo{pages}{28} (\bibinfo{year}{2019}).
	
	\bibitem[{\citenamefont{\mbox{Andrade}
	  et~al.}(2009)\citenamefont{\mbox{Andrade}, Oliveira, Moreira, and
	  Herrmann}}]{Andrade2009}
	\bibinfo{author}{\bibfnamefont{J.~S.} \bibnamefont{\mbox{Andrade}}},
	  \bibinfo{author}{\bibfnamefont{E.~A.} \bibnamefont{Oliveira}},
	  \bibinfo{author}{\bibfnamefont{A.~A.} \bibnamefont{Moreira}},
	  \bibnamefont{and} \bibinfo{author}{\bibfnamefont{H.~J.}
	  \bibnamefont{Herrmann}}, \bibinfo{journal}{Phys. Rev. Lett.}
	  \textbf{\bibinfo{volume}{103}}, \bibinfo{pages}{225503}
	  (\bibinfo{year}{2009}).
	
	\bibitem[{\citenamefont{Oliveira et~al.}(2011)\citenamefont{Oliveira, Schrenk,
	  Ara\'{u}jo, Herrmann, and Andrade}}]{Oliveira2011}
	\bibinfo{author}{\bibfnamefont{E.~A.} \bibnamefont{Oliveira}},
	  \bibinfo{author}{\bibfnamefont{K.~J.} \bibnamefont{Schrenk}},
	  \bibinfo{author}{\bibfnamefont{N.~A.~M.} \bibnamefont{Ara\'{u}jo}},
	  \bibinfo{author}{\bibfnamefont{H.~J.} \bibnamefont{Herrmann}},
	  \bibnamefont{and} \bibinfo{author}{\bibfnamefont{J.~S.}
	  \bibnamefont{Andrade}}, \bibinfo{journal}{Phys. Rev. E}
	  \textbf{\bibinfo{volume}{83}}, \bibinfo{pages}{046113}
	  (\bibinfo{year}{2011}).
	
	\bibitem[{\citenamefont{Dijkstra}(1959)}]{Dijkstra1959}
	\bibinfo{author}{\bibfnamefont{E.~W.} \bibnamefont{Dijkstra}},
	  \bibinfo{journal}{Numer. Math.} \textbf{\bibinfo{volume}{1}},
	  \bibinfo{pages}{269} (\bibinfo{year}{1959}).
	
	\bibitem[{\citenamefont{Cieplak et~al.}(1994)\citenamefont{Cieplak, Maritan,
	  and Banavar}}]{Cieplak1994}
	\bibinfo{author}{\bibfnamefont{M.}~\bibnamefont{Cieplak}},
	  \bibinfo{author}{\bibfnamefont{A.}~\bibnamefont{Maritan}}, \bibnamefont{and}
	  \bibinfo{author}{\bibfnamefont{J.~R.} \bibnamefont{Banavar}},
	  \bibinfo{journal}{Phys. Rev. Lett.} \textbf{\bibinfo{volume}{72}},
	  \bibinfo{pages}{2320} (\bibinfo{year}{1994}), ISSN \bibinfo{issn}{00319007}.
	
	\bibitem[{\citenamefont{Cieplak et~al.}(1996)\citenamefont{Cieplak, Maritan,
	  and Banavar}}]{Cieplak1996}
	\bibinfo{author}{\bibfnamefont{M.}~\bibnamefont{Cieplak}},
	  \bibinfo{author}{\bibfnamefont{A.}~\bibnamefont{Maritan}}, \bibnamefont{and}
	  \bibinfo{author}{\bibfnamefont{J.~R.} \bibnamefont{Banavar}},
	  \bibinfo{journal}{Phys. Rev. Lett.} \textbf{\bibinfo{volume}{76}},
	  \bibinfo{pages}{3754} (\bibinfo{year}{1996}), ISSN \bibinfo{issn}{10797114}.
	
	\bibitem[{\citenamefont{Porto et~al.}(1997)\citenamefont{Porto, Havlin,
	  Schwarzer, and Bunde}}]{Porto1997}
	\bibinfo{author}{\bibfnamefont{M.}~\bibnamefont{Porto}},
	  \bibinfo{author}{\bibfnamefont{S.}~\bibnamefont{Havlin}},
	  \bibinfo{author}{\bibfnamefont{S.}~\bibnamefont{Schwarzer}},
	  \bibnamefont{and} \bibinfo{author}{\bibfnamefont{A.}~\bibnamefont{Bunde}},
	  \bibinfo{journal}{Phys. Rev. Lett.} \textbf{\bibinfo{volume}{79}},
	  \bibinfo{pages}{4060} (\bibinfo{year}{1997}).
	
	\bibitem[{\citenamefont{Porto et~al.}(1999)\citenamefont{Porto, Schwartz,
	  Havlin, and Bunde}}]{Porto1999}
	\bibinfo{author}{\bibfnamefont{M.}~\bibnamefont{Porto}},
	  \bibinfo{author}{\bibfnamefont{N.}~\bibnamefont{Schwartz}},
	  \bibinfo{author}{\bibfnamefont{S.}~\bibnamefont{Havlin}}, \bibnamefont{and}
	  \bibinfo{author}{\bibfnamefont{A.}~\bibnamefont{Bunde}},
	  \bibinfo{journal}{Phys. Rev. E} \textbf{\bibinfo{volume}{60}},
	  \bibinfo{pages}{R2448} (\bibinfo{year}{1999}).
	
	\bibitem[{\citenamefont{Dobrin and Duxbury}(2001)}]{Dobrin2001}
	\bibinfo{author}{\bibfnamefont{R.}~\bibnamefont{Dobrin}} \bibnamefont{and}
	  \bibinfo{author}{\bibfnamefont{P.~M.} \bibnamefont{Duxbury}},
	  \bibinfo{journal}{Phys. Rev. Lett.} \textbf{\bibinfo{volume}{86}},
	  \bibinfo{pages}{5076} (\bibinfo{year}{2001}), ISSN \bibinfo{issn}{00319007}.
	
	\bibitem[{\citenamefont{Schrenk et~al.}(2012)\citenamefont{Schrenk, Ara\'{u}jo,
	  \mbox{Andrade}, and Herrmann}}]{Schrenk2012}
	\bibinfo{author}{\bibfnamefont{K.~J.} \bibnamefont{Schrenk}},
	  \bibinfo{author}{\bibfnamefont{N.~A.~M.} \bibnamefont{Ara\'{u}jo}},
	  \bibinfo{author}{\bibfnamefont{J.~S.} \bibnamefont{\mbox{Andrade}}},
	  \bibnamefont{and} \bibinfo{author}{\bibfnamefont{H.~J.}
	  \bibnamefont{Herrmann}}, \bibinfo{journal}{Sci. Rep.}
	  \textbf{\bibinfo{volume}{2}}, \bibinfo{pages}{348} (\bibinfo{year}{2012}).
	
	
	\bibitem[{\citenamefont{Schwartz et~al.}(2002)\citenamefont{Schwartz, Cohen,
	   \mbox{ben-Avraham}, Barab{\'a}si, and Havlin}}]{Schwartz2002}
	 \bibinfo{author}{\bibfnamefont{N.}~\bibnamefont{Schwartz}},
	   \bibinfo{author}{\bibfnamefont{R.}~\bibnamefont{Cohen}},
	   \bibinfo{author}{\bibfnamefont{D.}~\bibnamefont{\mbox{ben-Avraham}}},
	   \bibinfo{author}{\bibfnamefont{A.-L.}~\bibnamefont{Barab{\'a}si}},
	   \bibnamefont{and} \bibinfo{author}{\bibfnamefont{S.}~\bibnamefont{Havlin}},
	   \bibinfo{journal}{Phys. Rev. E} \textbf{\bibinfo{volume}{66}},
	   \bibinfo{pages}{015104(R)} (\bibinfo{year}{2002}).
	
	\bibitem[{\citenamefont{Buzna et~al.}(2006)\citenamefont{Buzna, Peters, and
	  Helbing}}]{Buzna2006}
	\bibinfo{author}{\bibfnamefont{L.}~\bibnamefont{Buzna}},
	  \bibinfo{author}{\bibfnamefont{K.}~\bibnamefont{Peters}}, \bibnamefont{and}
	  \bibinfo{author}{\bibfnamefont{D.}~\bibnamefont{Helbing}},
	  \bibinfo{journal}{Physica A} \textbf{\bibinfo{volume}{363}},
	  \bibinfo{pages}{132} (\bibinfo{year}{2006}).
	
	\bibitem[{\citenamefont{S\'anchez et~al.}(2002)\citenamefont{S\'anchez,
	  L\'opez, and Rodr\'iguez}}]{Sanchez2002}
	\bibinfo{author}{\bibfnamefont{A.~D.} \bibnamefont{S\'anchez}},
	  \bibinfo{author}{\bibfnamefont{J.~M.} \bibnamefont{L\'opez}},
	  \bibnamefont{and} \bibinfo{author}{\bibfnamefont{M.~A.}
	  \bibnamefont{Rodr\'iguez}}, \bibinfo{journal}{Phys. Rev. Lett.}
	  \textbf{\bibinfo{volume}{88}}, \bibinfo{pages}{048701}
	  (\bibinfo{year}{2002}).
	
	\bibitem[{\citenamefont{de~Noronha et~al.}(2018)\citenamefont{de~Noronha,
	  Moreira, Vieira, Herrmann, {Andrade}, and Carmona}}]{Noronha2018}
	\bibinfo{author}{\bibfnamefont{A.~W.~T.} \bibnamefont{de~Noronha}},
	  \bibinfo{author}{\bibfnamefont{A.~A.} \bibnamefont{Moreira}},
	  \bibinfo{author}{\bibfnamefont{A.~P.} \bibnamefont{Vieira}},
	  \bibinfo{author}{\bibfnamefont{H.~J.} \bibnamefont{Herrmann}},
	  \bibinfo{author}{\bibfnamefont{J.~S.} \bibnamefont{{Andrade}}},
	  \bibnamefont{and} \bibinfo{author}{\bibfnamefont{H.~A.}
	  \bibnamefont{Carmona}}, \bibinfo{journal}{Phys. Rev. E}
	  \textbf{\bibinfo{volume}{98}}, \bibinfo{pages}{062116}
	  (\bibinfo{year}{2018}).
	
	\bibitem[{\citenamefont{\mbox{de Gennes}}(1979)}]{deGennes1979}
	\bibinfo{author}{\bibfnamefont{P.-G.} \bibnamefont{\mbox{de Gennes}}},
	  \emph{\bibinfo{title}{Scaling concepts in polymer physics}}
	  (\bibinfo{publisher}{Cornell University Press}, \bibinfo{address}{Ithaca, New
	  York}, \bibinfo{year}{1979}).
	
	\bibitem[{\citenamefont{Chang and Aharony}(1991)}]{Chang1991}
	\bibinfo{author}{\bibfnamefont{I.}~\bibnamefont{Chang}} \bibnamefont{and}
	  \bibinfo{author}{\bibfnamefont{A.}~\bibnamefont{Aharony}},
	  \bibinfo{journal}{J. Phys. I} \textbf{\bibinfo{volume}{1}},
	  \bibinfo{pages}{313} (\bibinfo{year}{1991}).
	
	\bibitem[{\citenamefont{Poole et~al.}(1989)\citenamefont{Poole, Coniglio, Jan,
	  and Stanley}}]{Poole1989}
	\bibinfo{author}{\bibfnamefont{P.~H.} \bibnamefont{Poole}},
	  \bibinfo{author}{\bibfnamefont{A.}~\bibnamefont{Coniglio}},
	  \bibinfo{author}{\bibfnamefont{N.}~\bibnamefont{Jan}}, \bibnamefont{and}
	  \bibinfo{author}{\bibfnamefont{H.~E.} \bibnamefont{Stanley}},
	  \bibinfo{journal}{Phys. Rev. B} \textbf{\bibinfo{volume}{39}},
	  \bibinfo{pages}{495} (\bibinfo{year}{1989}).
	
	\bibitem[{\citenamefont{{OpenStreetMap contributors}}(2019)}]{OpenStreetMap}
	\bibinfo{author}{\bibnamefont{{OpenStreetMap contributors}}},
	  \emph{\bibinfo{title}{{Planet dump retrieved from https://planet.osm.org}}},
	  \bibinfo{howpublished}{\url{ https://www.openstreetmap.org}}
	  (\bibinfo{year}{2019}).
	
	\bibitem[{\citenamefont{{Google}}(2019)}]{GoogleDirections}
	\bibinfo{author}{\bibnamefont{{Google}}}, \emph{\bibinfo{title}{{Google
	  Directions API}}}, \bibinfo{howpublished}{\url{
	  https://developers.google.com/maps/documentation/directions/start}}
	  (\bibinfo{year}{2019}).
	
\end{thebibliography}

\end{document}